\documentclass[a4paper,11pt]{article}
\usepackage{jheppub} 
\usepackage[T1]{fontenc} 

\usepackage{xfrac}

\usepackage[makeroom]{cancel}

\def\be{\begin{equation}}
\def\ee{\end{equation}}
\def\bea{\begin{eqnarray}}
\def\eea{\end{eqnarray}}
\def\beq{\begin{eqnarray}}
\def\eeq{\end{eqnarray}}
\def\bas{\begin{subequations}\begin{eqnarray}}
\def\eas{\end{eqnarray}\end{subequations}}
\def\nn{\nonumber}

\def\eps{\varepsilon}

\def\tr{\text{tr}}

\def\ra{\rangle}

\def\f{\frac}

\def\SU{\text{SU}}
\def\su{\text{su}}
\def\SO{\text{SO}}

\def\SL{\text{SL}}
\def\su{\mathfrak{su}}
\def\so{\mathfrak{so}}
\def\sl{\mathfrak{sl}}

\newcommand{\C}{{\mathbb C}}
\newcommand{\N}{{\mathbb N}}
\newcommand{\R}{{\mathbb R}}
\newcommand{\Z}{{\mathbb Z}}

\newcommand{\cB}{{\mathcal B}}
\newcommand{\cE}{{\mathcal E}}

\newcommand{\cG}{{\mathcal G}}

\newcommand{\cJ}{{\mathcal J}}

\newcommand{\cH}{{\mathcal H}}
\newcommand{\cM}{{\mathcal M}}

\newcommand{\cO}{{\mathcal O}}

\newcommand{\cV}{{\mathcal V}}
\newcommand{\cD}{{\mathcal D}}
\newcommand{\cC}{{\mathcal C}}

\newcommand{\mat} [2] {\left ( \begin{array}{#1}#2\end{array} \right ) }

\def\vj{\vec{\j}}
\def\veta{\vec{\eta}}
\def\bz{\bar{z}}
\def\vtau{\vec{\tau}}
\def\pp{\partial}
\def\rd{\textrm{d}}

\def\ka{\kappa}

\def\eps{\epsilon}

\newcommand{\id}{\mathbb{I}}

\def\vJ{\vec{J}}

\def\vtau{\vec{\tau}}

\def\ra{\rangle}

\newcommand{\bes}{\begin{eqnarray}}
\newcommand{\ees}{\end{eqnarray}}

\renewcommand{\sl}{{\mathfrak{sl}}}

\newcommand{\qq}{\,,\quad}

\def\nn{\nonumber}
\def\pp{\partial}

\def\ka{\kappa}

\def\eps{\epsilon}

\def\vcJ{\vec{\cJ}}
\def\vJ{\vec{J}}
\def\vK{\vec{K}}
\def\vL{\vec{L}}

\def\bz{\bar{z}}

\def\hv{\widehat{v}}

\def\hJ{\widehat{J}}
\def\hK{\widehat{K}}
\def\hL{\widehat{L}}
\def\hE{\widehat{E}}
\def\hA{\widehat{A}}
\def\what{\widehat}

\def\hpi{\widehat{\pi}_{\phi}}

\def\hC{\hat{\mathfrak{C}}}

\def\hcE{\widehat{\cE}}
\def\hcC{\widehat{\cC}}
\def\hcH{\widehat{\cH}}

\def\hil{\mathfrak{H}}

\def\tt{\tilde{t}}
\def\dd{\mathrm{d}}

\def\pip{\pi_{\phi}}

\def\bw{\bar{w}}
\def\bW{\overline{W}}
\def\Heff{H_{\textrm{eff}}}
\def\hcC{\widehat{\cC}}
\def\hcH{\widehat{\cH}}

\def\hcH{\widehat{\cH}}
\def\hcC{\widehat{\cC}}

\title{\boldmath Conformal structure of FLRW Cosmology: \\
Spinorial representation and the $\so(3,2)$ algebra of observables}

 \author[a]{Jibril Ben Achour,}
\author[b]{Etera R. Livine,}

\affiliation[a]{Center for Gravitational Physics, Yukawa Institute for Theoretical Physics, Kyoto University, 606-8502, Kyoto, Japan}
\affiliation[b]{ENS de Lyon, Universit\'e Lyon 1, CNRS, 
Laboratoire de Physique, F-69342 Lyon, France}

\abstract{
It was recently shown that the homogeneous and isotropic cosmology of a massless scalar field coupled to general relativity exhibits a new hidden conformal invariance under Mobius transformation of the proper time, additionally to the invariance under time-reparamterization. The resulting Noether charges form a $\sl(2,\R)$ Lie algebra, which encapsulates the whole kinematics and dynamics of the geometry.
This allows to map FLRW cosmology onto conformal mechanics and formulate quantum cosmology in $\text{CFT}_1$ terms.
Here, we show that this conformal structure is embedded in a larger $\so(3,2)$ algebra of observables, which allows to present all the Dirac observables for the whole gravity plus matter sectors in a unified picture.
Not only this allows one to quantize the system and its whole algebra of observables as a single irreducible representation of $\so(3,2)$, but this also gives access to a scalar field operator $\hat{\phi}$ opening the door to the inclusion of non-trivial potentials for the scalar field. We expect this extended conformal structure to simplify the investigation of the quantum dynamics of inflationary cosmological backgrounds based on effective approaches to quantum cosmology.
}

\begin{document} 
\maketitle
\flushbottom

\newpage


\section*{Introduction}

Symmetries play a crucial role in theoretical physics, where they not only fix quantization ambiguities and constrain quantum correlators, but often entirely define the theory (up to dualities), especially in the context of quantum gravity. Previous works \cite{BenAchour:2019ufa, BenAchour:2019ywl,BenAchour:2018jwq, BenAchour:2017qpb} highlighted a hidden conformal invariance of the simplest gravitational system, the Friedman-Lema\^itre-Robertson-Walker (FLRW) cosmology of homogeneous and isotropic gravity coupled to a massless free scalar field\footnote{See also \cite{Pioline:2002qz} concerning similar structure in relation to the BKL conjecture.}. This conformal invariance of the cosmological action under Mobius transformation in proper time, is distinct from the time reparametrization invariance, which is the remnant of the diffeomorphism invariance of general relativity in this cosmological setting. While previous works in classical and quantum cosmology have also noticed the existence of a conformal structure, such as \cite{Pioline:2002qz, Bojowald:2007gc, Bojowald:2007bg, Erdmenger:2007xs, Baytas:2018qty}, the realization of the conformal symmetry discussed in \cite{BenAchour:2019ufa} is quite different.

At the classical level, this new symmetry leads by Noether theorem to conserved charges, encoding the evolution in proper time of the Hamiltonian, the extrinsic curvature and the volume \cite{BenAchour:2019ufa}. These charges form a $\sl(2,\R)$ Lie algebra, called the CVH algebra, a structure which was initially discussed in \cite{BenAchour:2018jwq, BenAchour:2017qpb}. This conformal invariance allows one to reformulate the phase space of the gravitational sector as a AdS${}_{2}$ space carrying the natural action of the 1D conformal group $\SL(2,\R)\sim\SU(1,1)$. Classically, this gravitational system can then be considered as a $\text{CFT}_1$ and map to the well known conformal mechanics of de Alfaro, Fubini and Furlan \cite{deAlfaro:1976vlx}, opening thus new directions for its quantization. Indeed, at the quantum level, the conformal invariance is a powerful tool, that fixes quantization ambiguities and the operator ordering. Moreover, importing the techniques developed for the quantization of conformal mechanics \cite{Chamon:2011xk, Jackiw:2012ur}, one can use this conformal structure to bootstrap the cosmological correlation functions \cite{BenAchour:2019ufa}. 

In this work, we show that this cosmological system enjoys an even larger conformal structure corresponding to an $\so(3,2)$ algebra which encodes the observables of the system. This refined conformal structure reveals itself when constructing and representing (Dirac) observables coupling the scalar field to the geometrical variables. Hence, contrary to the CVH structure discussed in previous works, it allows one to account for the whole phase space of the FLRW cosmology sourced by a scalar field and not only of the gravitational sector at fixed scalar momentum.
Not only this provides a more complete description of the model than the previous works focusing on the $\sl(2,\R)$ structure, but it will also allow to introduce new terms in the Hamiltonian which depend explicitly on the scalar field, that is include a non-trivial self-interacting potential. This step is crucial in order to extend the results of \cite{BenAchour:2019ufa} to more realistic cosmological systems such as inflationary backgrounds.

At the quantum level, the existence of this extended conformal structure implies that the theory can be quantized as a single irreducible $\SO(3,2)$ representation, which can be decomposed as a ladder of irreducible $\SL(2,\R)$ representations, each of them corresponding to a fixed value of the scalar field momentum. This Hilbert space of quantum states for the coupled geometry plus matter system carries a representation of the Dirac observables as quantum operators not only generating the evolution along the cosmological trajectories but also generating flows between those cosmological trajectories, thereby allowing to explore the whole cosmological phase space.


In a first section \ref{overview}, we present a short, self-contained, overview of the $\sl(2,\R)$ Lie algebra formed by the CVH observables and the corresponding conformal invariance of FLRW cosmology. The second section \ref{spinorial} introduces a parametrization of the cosmological phase space in terms of a complex 2-vector -or spinor. We show how to generate a $\so(3,2)$ algebra of observables from quadratic polynomials in the spinor components. The third section \ref{sec:so32} writes those $\so(3,2)$ observables explicitly in terms of the standard variables, the volume and scalar field and their conjugate momenta. In particular, we identify the subalgebra of Dirac observables. The fourth and final section \ref{su11HO} is dedicated to the quantization of the theory as a canonical quantization scheme of the spinor components. This preserves the $\so(3,2)$ algebra, and thus in particular the conformal invariance under $\SL(2, \R)$ transformations, at the quantum level. 
We conclude with a brief outlook towards the application of this new $\so(3,2)$ toolbox to quantum cosmology.

\section{The Conformal Symmetry of Cosmology} 
\label{overview}


We start with a quick review of the FLRW cosmology of general relativity coupled to a homogeneous and isotropic massless free scalar field, focusing on the $\sl(2,\R)$ framework and related conformal invariance introduced in the previous work \cite{BenAchour:2019ufa}. See also \cite{BenAchour:2019ywl, BenAchour:2018jwq, BenAchour:2017qpb} for details.

\subsection{The  $\sl(2,\R)$ structure of vacuum  cosmology}

Let us consider gravity minimally coupled to a massless scalar field $\phi$. 
The flat FLRW mini-superspace model is defined by focusing on homogeneous isotropic metrics, given by the ansatz
\be
\rd s^2 =-N(t)^2 \rd t^2 +a(t)^2\delta_{ij}\rd x^i\rd x^j
\,,
\ee
in terms of the lapse function $N(t)$ and the scalar factor $a(t)$. We similarly assume that the scalar field $\phi$ is homogeneous and only depends on the time $t$.
Assuming a vanishing cosmological constant $\Lambda=0$, the reduced FRW cosmological action is given by the integration of the Einstein-Hilbert action over a fiducial 3D cell of volume $V_{o}$ and reads:
\be
\label{FRWaction}
S[a,N,\phi]
\,=\,
V_{o}\int \rd t\left[
-\f{3}{8\pi G}\f{a\dot{a}^2}{N}+\f{a^3}{2N}\dot\phi^2
\right]
\,,
\ee
with the Newton constant $G$.
The Hamiltonian analysis of this action defines the conjugate momenta to the scale factor and scalar field:
\be
\pi_{a}=-\f{3V_{o}}{4\pi G}\f{a\dot a}{N}
\,,\qquad
\pi_{\phi}=\f{a^3V_{o}}{N}\dot\phi
\,,
\ee
and writes the action as a fully constrained system,
\be
S=\int\rd t\,\big{[}
\dot a \pi_{a}+\dot \phi \pi_{\phi}-N\cH
\big{]}
\,,
\ee
where the lapse $N$ plays the role of a Lagrange multiplier and the Hamiltonian constraint $\cH$ balances the energy of the matter field with the energy of the geometry:
\be
\cH=\f1{2V_{o}}\left(
\f{\pi_{\phi}^2}{a^3}-\f{4\pi G}3 \f{\pi_{a}^2}a
\right)
\,.
\ee
It is convenient to x introduce a volume variable, absorbing the volume of the fiducial cell. The  canonical transformation reads:
\be
v=a^3 V_{o}
\,,\qquad
b=-\f1{3V_{o}}\f{\pi_{a}}{a^2}=\f{1}{4\pi G}\f{\dot a}{Na}
\,.
\ee
Then the phase space of  homogeneous, isotropic and flat FRW cosmology is given by canonical variables,
\be
\{ b, v \} = 1
\,,\qquad
\{ \phi, \pi_{\phi}\} = 1
\,,
\ee
provided with the Hamiltonian constraint:
\be
\cH
= \f12\left(
\f{\pi_{\phi}^2}v-\ka^2 vb^2
\right)
=
\f{\pi_{\phi}^2}{2v}+\cH_{g}
\,\simeq0\,,
\ee
where the constant $\ka=\sqrt{12\pi G}$ is proportional to the Planck length and encodes the coupling between matter and geometry.

\medskip

Since the scalar field has no potential or mass,  its momentum $\pi_{\phi}$ is a constant of motion,
\be
\{\cH,\pi_{\phi}\}=0
\,.
\ee
Let us start by describing the theory for $\pi_{\phi}=0$, i.e. let us set the scalar field to 0 and focus on the purely gravitational sector.
The FRW phase space for pure gravity is given by the canonical pair, $\{ b, v\} = 1$, provided with the Hamiltonian constraint $\cH_g$,
\be
\cH_g =  - \frac{\ka^2}{2} v b^2 = 0
\,.
\ee
%
%
We further introduce the integrated trace of the extrinsic curvature,
\be
\int_{V_o} \rd^3x \, \sqrt{|\gamma|} K
=
\int_{V_o} \rd^3x \, a^3\f{3\dot a}{Na}
=
\ka^2 bv
\,.
\ee 
We recognize this quantity $\cC =bv$ as the generator of dilatations of for the geometry canonical pair:
\be
e^{\{ \eta  \cC, \cdot\}} \, b = e^{-\eta} b
\,,
\qquad
e^{\{ \eta  \cC, \cdot\}} \, v = e^{+\eta} v
\,.
\ee 
%
%
Since the volume $v$ and the Hamiltonian constraint $\cH_{g}$ are respectively  of dimension $+1$ and $-1$ with respect to scale transformations, it turns out that, together with the dilatation generator $\cC$, they form a closed $\sl(2,\R)$ Lie algebra:
\be
\{ \cC, v \} = v
\,, \quad
\{ \cC, \cH_g \} = - \cH_{g}
\,, \quad
 \{ v, \cH_g \} = \ka^{2}\cC
 \,,
\ee
which is referred to as the CVH algebra \cite{BenAchour:2017qpb,BenAchour:2018jwq,BenAchour:2019ywl}.
To set it in the standard $\sl(2,\R)\sim\su(1,1)$ basis,
\be
\{ j_z, k_x\} = k_y
\,, \quad
\{ j_z, k_y\} = - k_x
\,, \quad
\{ k_x, k_y\} = - j_z 
\,,
\nn
\ee
the change of basis between the three cosmological observables and the $\sl(2,\R)$ generators reads:
\be
 \cC = k_y 
 \,, \quad
 v = \sigma\ka^{3}( k_x + j_z )
 \,, \quad
 \cH_g =  \frac{1}{2\sigma\ka} \big{(} k_x - j_z \big{)}
 \,,
\ee
\be
j_z =
\f v{2\sigma\ka^{3}}-\sigma\ka\cH_g
\,, \quad
k_x =
\f v{2\sigma\ka^{3}}+\sigma\ka\cH_g
\,, \quad
k_y = \cC
\,,
\ee
where $\sigma$ is an arbitrary dimensionless real parameter.
While the dilatation $\cC$ is a pure boost generator, the volume $v$ and the Hamiltonian constraint $\cH_{g}$ are null vectors
in the Lie algebra $\sl(2,\R)$. 
%
%
%
This isomorphism between the CVH algebra and the $\sl(2,\R)$ Lie algebra leads to a vanishing $\sl(2,\R)$-Casimir:
\be
\label{gravCasimir}
\mathfrak{c}_{\mathfrak{sl}(2,\R)}
= j^2_z - k^2_x - k^2_y
= -2v\cH_{g}-\cC^{2}
= 0
\,.
\ee
This implies that vacuum homogeneous and isotropic general relativity can be described at the quantum level by a null representation of $\sl(2,\R)$, as advocated already in \cite{BenAchour:2017qpb} and shown in details in \cite{BenAchour:2019ywl}.

As explained in \cite{BenAchour:2017qpb} and reviewed below, the inclusion of matter preserves the $\sl(2,\R)$ structure and simply induces a shift in the Casimir. Furthermore, it was understood in \cite{BenAchour:2019ufa} that this $\sl(2,\R)$ structure descends from a conformal invariance of the  action for gravity plus scalar matter allowing to map it the cosmological system onto the conformal quantum mechanics introduced by de Alfaro, Fubini and Furlan \cite{deAlfaro:1976vlx}.

\subsection{Conformal symmetry for FRW cosmology}

Coming back to the full theory with gravity coupled to a scalar field by the Hamiltonian constraint $\cH$, it turns out that the CVH algebra and its mapping to $\sl(2,\R)$ is preserved, with the same identification:
\be
\label{CVHgm}
 \cC = K_y 
 \,,\quad
 v = \sigma\ka^{3}( K_x + J_z )
 \,,\quad
 \cH =  \frac{1}{2\sigma\ka} \big{(} K_x - J_z \big{)}
 \,,
\ee
\be
\{ J_z, K_x\} = K_y
\,, \quad
\{ J_z, K_y\} = - K_x
\,, \quad
\{ K_x, K_y\} = - J_z 
\,.
\nn
\ee
The full Hamiltonian constraint $\cH$ is still a null generator\footnotemark{} in $\sl(2,\R)$.
\footnotetext{
If we take into account a non-vanishing cosmological constant $\Lambda\ne 0$, the Hamiltonian constraint  becomes a time-like or space-like $\su(1,1)$ generator depending on the sign of the cosmological constant.
}
And notice that the dilatation generator $\cC=K_{y}=k_{y}$ remains unchanged: it only acts on the geometrical sector and does not affect the scalar field.

The main effect of the inclusion of matter is a shift of the Casimir. Indeed, the $\sl(2,\R)$-Casimir does not vanish anymore, it is now strictly negative and is given by the scalar field density energy:
\be
\label{Csl2R}
\mathfrak{C}_{\mathfrak{sl}(2,\R)} 
=   J^2_z - K^2_x - K^2_y
= - 2v \cH - \cC^2  
=-\frac{\pi^2_{\phi}}{\ka^2}
\,.
\ee
This means that the system gravity plus matter will be described at the quantum level  by a space-like representation of $\SL(2,\R)$, which depends on the value of the scalar field momentum $\pi_{\phi}$.

\medskip

An elegant way to derive the matter coupling from the pure gravity sector is to derive the $\sl(2,R)$ generators $J_{z},K_{x},K_{y}$ for gravity plus matter from the  $\sl(2,R)$ generators  $j_{z},k_{x},k_{y}$ for pure gravity using a non-linear map.
Indeed, we can directly extend the $\sl(2,\R)$ structure of the pure gravity case while preserving the $\sl(2,\R)$ Poisson brackets, by keeping the boost in the $y$-direction, $K_{y}=k_{y}$, and modifying the generators in the $x$ and $z$ directions:
\be
J_{z}=j_{z}+\f\lambda{2(j_{z}+k_{x})}
\qq
K_{x}=k_{x}-\f\lambda{2(j_{z}+k_{x})}
\,,
\ee
where $\lambda$ is a constant, i.e. it is a central element, which Poisson-commutes with the $\sl(2,\R)$ generators $j_{z},k_{x},k_{y}$.
This works because $(k_{x}+j_{z})^{-1}$ and $(k_{x}-j_{z})$ scale the same way under dilatations generated by $k_{y}$.
It is straightforward to check that these satisfy the correct $\sl(2,\R)$ Poisson brackets, while shifting the Casimir by $\lambda$:
\be
\mathfrak{C}
=
J^2_z - K^2_x - K^2_y 
=
j^2_z - k^2_x - k^2_y
+\lambda
=
\mathfrak{c}
+\lambda
\,.
\ee
These extended $\su(1,1)$ generators exactly fit the definition of the CVH algebra \eqref{CVHgm} for gravity plus scalar matter if $\lambda$ is fixed by the scalar field momentum:
\be
\lambda =  -\frac{\pi^2_{\phi}}{\ka^2}\,.
\ee
In the context of conformal quantum mechanics, the  same mapping exists between the free particle and  conformal mechanics \cite{deAlfaro:1976vlx}.

\medskip

This CVH algebra encodes the dynamics of the theory. 
Indeed, it describes the iterative Poisson brackets of the volume with the Hamiltonian constraint, which gives the evolution of the volume in proper time $\tau$, related to the time coordinate by the lapse factor $\rd \tau =N\rd t$,
\be
\f{\rd v}{\rd \tau}=\f1N\f{\rd v}{\rd t}
=
\{v,\cH\}
=
\ka^2 \cC\,,
\ee
\be
\f{\rd \cC}{\rd \tau}
=
\{\cC,\cH\}
=
-\cH
\,,
\qquad
\f{\rd \cH}{\rd \tau}
=
\{\cH,\cH\}
=0\,.
\ee
These equations of motion are straightforward to integrate:
\be
\label{cosmotraj}
\left|\begin{array}{ccl}
\cH(\tau)
&=&
\cH^{(0)}\,,
\vspace*{1mm}\\
\cC(\tau)
&=&
\cC^{(0)}-\tau \cH^{(0)}\,,
\vspace*{1mm}\\
v(\tau)
&=&
v^{(0)}+\tau\ka^{2}\cC^{(0)}-\f{\tau^{2}}2\ka^{2}\cH^{(0)}\,.
\end{array}\right.
\ee
We can invert this trajectory to express the initial conditions as constants of motion:
\be
\left|\begin{array}{ccl}
\cH^{(0)}
&=&
\cH\,,
\vspace*{1mm}\\
\cC^{(0)}
&=&
\cC+\tau \cH\,,
\vspace*{1mm}\\
v^{(0)}
&=&
v-\tau\ka^{2}\cC-\f{\tau^{2}}2\ka^{2}\cH
\,.
\end{array}\right.
\ee
As shown in \cite{BenAchour:2019ufa}, these three constants of motions turn out to be the three conserved Noether charges associated to the invariance of the reduced cosmological action \eqref{FRWaction} under Mobius transformations of the proper time:
\beq
\label{Mobius}
\tau&\mapsto& \tilde{\tau}=\f{\alpha\tau+\beta}{\gamma\tau+\delta}\,, \\
v & \mapsto& \tilde{v}(\tilde{\tau})=\f{v(\tau)}{(\gamma\tau+\delta)^{2}}\,, \nn\\
\phi &\mapsto& \tilde{\phi}(\tilde{\tau})=\phi(\tau)\,,\nn
\eeq
with $\alpha\delta-\beta\gamma=1$ and $\alpha,\beta,\gamma,\delta$ real. As such, the CVH algebra reflects the conformal invariance of FRW cosmology under the one-dimensional conformal group $\SL(2,\R)$.

\medskip

One shortcoming that we would like to address in the present paper is that this formalism does not does not allow us to straightforwardly integrate the equation of motion for the scalar field, in particular it does not provide us with a constant of motion involving $\phi$.
As a matter of fact, the procedure above can not give  the evolution of the scalar field and we have to integrate its equation of motion by hand,
\be
\rd_{\tau}\phi=\{\phi,\cH\}=\f{\pi_{\phi}}{v}
=\f{\pi_{\phi}}{v^{(0)}+\tau\ka^{2}\cC^{(0)}-\f{\tau^{2}}2\ka^{2}\cH^{(0)}}
\,.\nn
\ee
Assuming that we are on-shell, the Hamiltonian constraint always vanishes, $\cH^{(0)}=0$. Furthermore the dilatation generator is a weak Dirac observable, $\{\cH,\cC\}=\cH\simeq0$, thus a constant of motion, $\cC(\tau)=\cC^{(0)}$. Solving the Hamiltonian constraint implies that $\ka^{2}\cC^{2}=\pi_{\phi}^{2}$, with two branches of solution:
\be
\cC=\pm\f{\pi_{\phi}}{\ka}\,.
\ee
Since $\rd_{\tau}v=\ka^2 \cC$, this gives the speed of the volume:
\be
v= v^{(0)}+\tau \ka^{2}\cC^{(0)}= v^{(0)}\pm\ka\pi_{\phi}\tau\,,
\ee
where the sign decides if the universe are in a contracting or expanding phase.
Plugging this linear growth of the volume in the equation of motion for the scalar field gives:
\be
\rd_{\tau}\phi
=
\f{\pm\ka\cC^{(0)}}{v^{(0)}+\tau\ka^{2}\cC^{(0)}}
\,\,\Rightarrow\,\,
\phi(\tau)=\phi^{(0)}\pm\f1\ka\ln v
\,.
\ee
This is usually written in a deparametrized fashion by getting rid of the time variable (see appendix \ref{deparametrization} for more details), reflecting the invariance of the theory under time reparametrization:
\be
\label{vphitraj}
v=v^{(0)}\,e^{\pm\ka(\phi-\phi^{(0)})}
\,.
\ee
This describes the cosmological evolution in terms of the scalar field clock $\phi$. We have two regimes\footnotemark, with contracting or expanding trajectories, determined by the sign $\pm$.
\footnotetext{
In modified cosmological models with a big bounce, such loop quantum cosmology and related approaches, these two regimes fuse and contracting trajectories bounce back into expanding trajectories. The $\sl(2,\R)$ formalism, with the CVH algebra and conformal invariance, can also be applied to these modified cosmological models \cite{BenAchour:2018jwq, BenAchour:2019ywl}.
}
In each phase, the on-shell relation between the volume and the scalar field leads to a different set of reduced phase space variables.
More precisely, the Hamiltonian constraint factorizes as
\be
\cH = \frac{\pi^2_{\phi}}{2 v} - \frac{\ka^2v b^2}{2} 
=\f1{2 v}(\pip-\ka\cC)(\pip+\ka\cC)
\simeq 0
\,,
\ee
with the two branches of solutions, $\pip\simeq\eps\cC$, with $\eps=\pm$, defining the two expanding and contracting phases.
The two pairs of variables, 
\be
\label{DiracVB}
\cV_{\eps}=v e^{- \eps\ka\phi}
\,,\quad
\cB_{\eps}=b e^{+ \eps\ka\phi}=\f{\cC}{\cV_{\eps}}
\,,
\ee
are respectively weak Dirac observables in the corresponding branch,
\be
\{\cV_{\eps},\cH\}
=
\f{\cV}{v}
\left[
\cC- \eps\frac{\pi_{\phi}}{\ka}
\right]
\simeq 0
\,,\quad
\{\cB_{\eps},\cH\}\simeq0
\,,
\ee
where the weak equality  holds on the appropriate branch, i.e. assuming $\pip\simeq\eps\ka\cC$. Let us emphasize that those Poisson brackets do not weakly vanish if we only assume that $\cH\simeq 0$. Thus, $\cV_{\eps}$ and $\cB_{\eps}$ can only be thought of as Dirac observables if we choose the $\eps$-branch and switch the original Hamiltonian constraint $\cH\simeq 0$ for the stronger constraint $\pip\simeq \eps\ka\cC$. We will see in the section \ref{Obs-so32} how to upgrade them to complete Dirac observables.

The point here is that these weak Dirac observables are  the evolving constants of motion, encoding the evolution of the volume $v$ and extrinsic curvature $b$ with respect to the scalar field $\phi$. See \cite{Rovelli:2001bz, Dittrich:2005kc} for details on the notion of evolving constant of motion.
They define a canonical pair of conjugate variables on the reduced phase space:
\be
\{\cB_{\eps},\cV_{\eps}\}=1\,.
\ee

\medskip

The objective of the present work is to extend the $\sl(2,\R)$ structure to remedy this asymmetry between the treatment of the gravitational and matter sectors and construct observables involving the scalar field $\phi$. Not only this will allow to represent the scalar field at the quantum level  as an operator acting on $\sl(2,\R)$ representations, but it will also open the door to the inclusion of a scalar field mass or potential implying terms in the Hamiltonian which explicitly depend on $\phi$.

\subsection{Mapping onto conformal mechanics}
\label{CQM}

We conclude this brief overview of the conformal structure of FRW cosmology by its mapping onto conformal mechanics, following \cite{Pioline:2002qz,BenAchour:2019ufa}.
Starting from the cosmological action \eqref{FRWaction} in terms of a Lagrangian depending on the scale factor $a$ and the scalar field $\phi$, we can absorb the lapse in the proper time $\rd \tau =N \rd t$ and re-write it in terms of the volume $v$:
\be
S_{FRW}[v(\tau),\phi(\tau)]
=
\int \rd\tau\,
\left[
-\f{(\rd_{\tau}v)^2}{2\ka^2v}+\f v2(\rd_{\tau}\phi)^2
\right]
\ee
The equation of motion for the scalar field implies that $v\rd_{\tau}\phi$ is constant, which we recognize as the scalar field momentum $\pip=v\rd_{\tau}\phi$. To plug this back in the action, and obtain an action solely on the volume $v$, we have a do a Legendre transform on the canonical pair $(\phi,\pip)$, which gives:
\beq
S_{FRW}[v(\tau)]
&=&
S_{FRW}[v(\tau),\phi(\tau)]
-
\int \rd\tau\,
\pip\rd_{\tau}\phi
\nn\\
&=&
\int \rd\tau\,
\left[
-\f{(\rd_{\tau}v)^2}{2\ka^2v}-\f{\pip^2}{2v}
\right]
\,,
\eeq
where $\pip$ is now simply a coupling constant.
Introduce the new variables:
\be
q
=
\f{2\sqrt{v}}\ka
=
a^{3/2}\sqrt{\f{V_{o}}{3\pi G}}
\,,
\ee
This allows to re-write the action for FRW cosmology as:
\be
S_{FRW}[q]=
-\f12\int \rd\tau\,\left[
(\rd_{\tau}q)^{2}-\f{g}{q^{2}}
\right]
\,.
\ee
with the coupling constant in front of the potential given by
\be
g
=
-\f{4\pi_{\phi}^{2}}{\ka^{2}}
=
-\f1{3\pi G}\left(V_{o} a^{3}\rd_{\tau}\phi\right)^{2}
\le 0
\,.
\ee
The action above is then exactly the action for the conformal particle \cite{deAlfaro:1976vlx}, which is also the action for the relative motion of the 2-body Calogero model.
The potential in $q^{-2}$ ensures the conformal invariance of the model and the homogeneous scaling of the Hamiltonian under scale transformations. The main difference between conformal mechanics and FRW cosmology is the lapse variable $N$, which implies the invariance under time reparametrization. Thus FRW cosmology appears as a special case of conformal mechanics for which the energy vanishes.

To get this, we either perform the canonical analysis of the action above or  the mapping to conformal mechanics directly from the Hamiltonian formulation of the action for FRW cosmology,
\be
S_{FRW}=\int \rd t\,\bigg{[}
\dot b v +\dot \phi \pi_{\phi} -\f N2\Big{(}
\f{\pi_{\phi}^{2}}v-\ka^{2}vb^{2}
\Big{)}
\bigg{]}
\,.
\ee
Taking into account that $\pip$ is constant, we can simply discard the scalar field  and focus on the dynamics of the volume. Then switching to the new variables  $q=2\ka^{-1}\sqrt{v}$ and its conjugate momentum $\pi_{q}=-\ka b \sqrt{v}$, the action reads:
\beq
S_{FRW}&=&
\int \rd t\,\bigg{[}
\dot q \pi_{q}
+\f N2\Big{(}
\pi_{q}^{2}-\f{4\pi_{\phi}^{2}}{\ka^{2}}\f1{q^{2}}
\Big{)}
\bigg{]}
\\
&=&
\int \rd \tau\,\bigg{[}
\pi_{q}\rd_{\tau}q
+\f 12\Big{(}
\pi_{q}^{2}-\f{4\pi_{\phi}^{2}}{\ka^{2}}\f1{q^{2}}
\Big{)}
\bigg{]}
\,,\nn
\eeq
which is the Hamiltonian form of  conformal mechanics, with the energy of the corresponding conformal particle exactly equal to the FRW Hamiltonian constraint, and thus vanishing on-shell:
\be
E_{CM}=\f12\left[
\f{\pi_{\phi}^{2}}v-\ka^{2}b^{2}v
\right]
=\cH\sim0
\,.\nn
\ee

\section{Spinorial Parametrization}
\label{spinorial}

Up to now, we have reformulated the cosmological phase space as a $\sl(2,\R)$ Lie algebra. However, we have seen that this does not allow access to the scalar field $\phi$. Indeed, the three $\sl(2,\R)$ generators, $J_{z}$, $K_{x}$ and $K_{y}$, only encode the information on the volume $v$, its conjugate momentum $b$, which measures the extrinsic curvature, and the scalar field momentum $\pi_{\phi}$. In this section, we will show that it is possible to extend the $\sl(2,\R)$ structure to include the scalar field $\phi$.
The main tool is to realize the $\sl(2,\R)$ Lie algebra in terms of a spinor, i.e. a complex 2-vector $z=(z^{0},z^{1})$, provided with the canonical Poisson bracket,
\be
\label{spinorPB}
\{z^{A},z^{B}\}=0
\,,\quad
\{z^{A},\bz^{B}\}=-i\delta^{AB}
\,.
\ee
This 4-dimensional phase space will now allow to faithfully encode all the dynamical variables of the original cosmological phase space, $(b,v)$ and $(\phi,\pi_{\phi})$.
More precisely, the scalar field moment $\pip$ will give the value of the $\sl(2,\R)$  Casimir while the scalar field  $\phi$, as its its conjugate variable, will generate shifts of this Casimir. At the quantum level, this implies that working at constant $\pip$ amounts to working in a fixed $\sl(2,\R)$ irreducible representation while acting with the scalar field operator $\hat{\phi}$ will switch between $\sl(2,\R)$ representations.

Moreover, the spinorial formulation also offers  improved features, which we will study in more details in the rest of the paper:
\begin{itemize}

\item It allows to describe the cosmological evolution as a $\SL(2,\R)$ flow in its fundamental representation. Indeed, $\SU(1,1)\sim\SL(2,\R)$ transformations simply act as 2$\times$2 complex matrices on the complex 2-vector $z$. Using this complex parametrization simplifies the description of the cosmological trajectories.

\item The spinor $z$ allows to extend the CVH algebra from a $\sl(2,\R)$ algebra to a $\so(3,2)$ Lie algebra of observables, enriching the algebraic structure  of the phase space.

\item The complex phase space can be canonically quantized in terms of a pair of harmonic oscillators, which allows to represent at the quantum level the whole tower of $\SL(2,\R)$ representations with negative quadratic Casimir $\mathfrak{C}<0$. This naturally fixes quantization ambiguities and operator ordering.

\end{itemize}



\medskip

Let us introduce a spinor $z=(z^{0},z^{1})\in\C^{2}$ provided with a canonical Poisson bracket:
\be
\{z^{A},z^{B}\}=0
\,,\quad
\{z^{A},\bz^{B}\}=-i\,\delta^{AB}
\,.
\ee
We can recover the $\sl(2,\R)$ Lie algebra as quadratic polynomials in the spinor,
\be
\begin{array}{ccl}
J_{z}&=&\f12\,\left[|z^{0}|^{2}-|z^{1}|^{2}\right]
\vspace*{1mm}
\\
K_{+}&=&\f12\,\left[(\bz^{0})^{2}-(z^{1})^{2}\right]
\vspace*{1mm}
\\
K_{-}
&=&\f12\,\left[(z^{0})^{2}-(\bz^{1})^{2}\right]
=\overline{K_{+}}
\end{array}
\ee
which leads back to the $\SL(2,\R)$ poisson brackets with $K_{\pm}=K_{x}\pm i K_{y}$,
\be
\{J_{z},K_{\pm}\}=\mp i K_{\pm}
\,,\quad
\{K_{+},K_{-}\}=2i J_{z}
\,.
\nn
\ee
We obtain a negative quadratic Casimir as wanted\footnotemark:
\be
\mathfrak{C}
=   J^2_z - K_{+}K_{-}
=-\f{1}4\Big{[}i(\bz^{0}\bz^{1}-z^{0}z^{1})\Big{]}^{2}
=-\cE^{2}\le 0\,.
\nn
\ee
\footnotetext{
This construction is different from the  spinorial representation of the $\sl(2,\R)\sim\su(1,1)$ algebra for positive or vanishing Casimir, $\mathfrak{C}\ge 0$, which leads to time-like unitary representations at the quantum level and which is used for instance to define $\SU(1,1)$ coherent states for deparametrized quantum cosmology \cite{Livine:2012mh}. 
}
The quantity $\cE$ is itself a Casimir of the $\sl(2,\R)$ algebra, i.e. it Poisson-commutes with the $\sl(2,\R)$ generators:
\be
\cE=\f i2(\bz^{0}\bz^{1}-z^{0}z^{1})
\,,\quad
\{J_{z},\cE\}=\{K_{x,y},\cE\}=0
\,.
\ee

\subsection{Cosmological trajectories as $\SL(2,\R)$ flow}
\label{su11trajectory}

To look at the exponentiated action of $\SL(2,\R)$ transformations, it is convenient to switch self-dual complex variables:
\be
w=\f1{\sqrt2}(z^{0}-\bz^{1})
\,,\quad
W=\f1{\sqrt2}(z^{0}+\bz^{1})
\,,
\ee
\be
\{w,W\}=0
\,,\quad
\{w,\bW\}=-i
\,,\quad
\{w,\bw\}=0\,.
\nn
\ee
This allows to write the Hermitian 2$\times$2 matrix of the $\sl(2,\R)$ Lie algebra,
\be
\label{M}
M=\mat{cc}{J_{z} & K_{-}\\ K_{+} & J_{z}}
\,,
\ee
as a complex scalar product,
\be
\cM\equiv \mat{c}{w\\ \bw}\,\mat{c}{W \\ \bW}^{\dagger}
\,,\quad
\tr(\cM\tau_{z})=2i\cE
\,,
\ee
\be
M=\cM-\f12\tr(\cM\tau_{z})\,\tau_{z}
\,,\qquad
\left|\begin{array}{ccl}
wW&=&K_{-}\,,\\
\bw\bW&=& K_{+}\,,\\
w\bW&=&J_{z}+i\cE\,.
\end{array}\right.
\nn
\ee
The map from the complex variables $w,W$ to the $\sl(2,\R)$ generators is not one-to-one, since it sends 4 real parameters onto 3 real variables. Indeed, the definition above of the generators $\vcJ=(J_{z},K_{x},K_{y})$ as quadratic polynomials in the spinor is invariant under  real rescalings:
\be
\left|\begin{array}{ccl}
w&\rightarrow&\lambda\,w\\
W&\rightarrow&\lambda^{-1}\,W
\end{array}\right.
\,,\quad \forall\lambda\in\R\,.
\ee
The advantage of this formulation is that  $\SL(2,\R)$ group elements now act on the spinor in the fundamental representation of $\SU(1,1)$. Let us introduce the Lorentzian Pauli matrices,
\be
\tau_{z}=\mat{cc}{1 & 0 \\ 0 & -1}
\,,\,\,
\tau_{x}=\mat{cc}{ 0 & 1 \\ -1 &0}
\,,\,\,
\tau_{y}=\mat{cc}{ 0 & -i \\-i &0}
\,.
\nn
\ee
These matrices square to the identity, $\tau_{z}^{2}=\id$ but $\tau_{x}^{2}=\tau_{y}^{2}=-\id$, and satisfy the $\sl(2,\R)\sim\su(1,1)$ commutation relations:
\be
[\tau_{z},\tau_{x}]=2i\tau_{y}
\,,\quad
[\tau_{x},\tau_{y}]=-2i\tau_{z}
\,,\quad
[\tau_{y},\tau_{z}]=2i\tau_{x}
\,.
\nn
\ee
Then, we can compute the Poisson brackets of the $\sl(2,\R)$ generators with the spinors:
\be
\left\{\veta\cdot\vcJ,\mat{c}{w\\\bw}\right\}
\,=\,
\f i2\veta\cdot\vtau\,\mat{c}{w\\\bw}
\,,
\ee
and similarly for $W$, with the  Lorentzian signature scalar product $\veta\cdot\vcJ=\eta_{z}J_{z}-\eta_{x}K_{x}-\eta_{y}K_{y}$.
This exponentiates to the group action:
\be
e^{\{\veta\cdot\vcJ,\cdot\}}\,\mat{c}{w\\\bw}\,=\,e^{\f12 \veta\cdot\vtau}\,\mat{c}{w\\\bw}
\,.
\ee
This leads to the exponentiated $\SL(2,R)$-action on the  matrix $M$ of generators as the conjugation by $\SU(1,1)$ matrices,
\be
\{\vcJ,M\}=\f i 2\,\big{(}
\vtau M -M \vtau^{\dagger}
\big{)}\,,
\ee
\be
\label{conjugation}
e^{\{\veta\cdot\vcJ,\cdot\}}M=GMG^{\dagger}
\quad\textrm{with}\,\,
G=e^{\f12 \veta\cdot\vtau}\in\SU(1,1)
\,.
\ee


\medskip

This allows to integrate the exponentiated flow of any $\sl(2,\R)$ generators in the CVH algebra. In particular, we can derive the Hamiltonian flow generated by the Hamiltonian constraint $\cH$ and recover the cosmological trajectories. Indeed, applying the formula above to  $e^{-\tau\{\cH,\cdot\}}$ with $\cH=(K_{x}-J_{z})/2\sigma\ka$ leads to into null $\SU(1,1)$ transformations:
\be
\cG_{\tau}=e^{-i\f \tau{4\sigma \ka}(\tau_{x}-\tau_{z})}=\id-i\f \tau{4\sigma \ka}(\tau_{x}-\tau_{z})
\,,
\ee
where $\tau_{x}-\tau_{z}$ is a nilpotent matrix with vanishing square. This gives the cosmological evolution in proper time of the $\sl(2,\R)$ generators:
\be
\left|\begin{array}{ccl}
J_{z}(\tau)
&=&
J_{z}^{(0)}+\f \tau{2\sigma \ka} K_{y}^{(0)}-\f {\tau^2}{8\sigma^2 \ka^2}(K_{x}^{(0)}-J_{z}^{(0)})
\,,
\vspace*{1mm}\\
K_{x}(\tau)
&=&
K_{x}^{(0)}+\f \tau{2\sigma \ka}K_{y}^{(0)}-\f {\tau^2}{8\sigma^2 \ka^2}(K_{x}^{(0)}-J_{z}^{(0)})
\,,
\vspace*{1mm}\\
K_{y}(\tau)
&=&
K_{y}^{(0)}-\f \tau{2\sigma \ka} (K_{x}^{(0)}-J_{z}^{(0)})
\,,
\end{array}\right.
\ee
which we can translate into the proper time evolution of the cosmological observables and recover the trajectories derived earlier in \eqref{cosmotraj} from the knowledge of constants of motion (identified as the Noether charges for the conformal symmetry):
\be
\left|\begin{array}{ccl}
v(t)
&=&
v^{(0)}+\tau\ka^2\cC^{(0)}-\f{\tau^{2}}2\ka^2\cH^{(0)}
\,,
\vspace*{1mm}\\
\cH(t)
&=&
\cH^{(0)}
\,,
\vspace*{1mm}\\
\cC(\tau)
&=&
\cC^{(0)}-\tau\cH^{(0)}
\,.
\end{array}\right.
\ee

\subsection{$\so(3,2)$ Lie algebra from the spinor}

The spinor phase space not only carries a representation of the $\sl(2,\R)\sim\su(1,1)$ Lie algebra but it also provides a representation of the much larger $\so(3,2)$ Lie algebra, which  can be thought of as the conformal group in 2+1 dimensions.
Let us first define the $\so(3,2)$ generators, then we will see in the next section that these generators correspond to cosmological observables involving the scalar field $\phi$. This extension of the CVH algebra thus allows to faithfully represent all the whole phase space of the FRW cosmology of a scalar field.

Let us  consider the whole set of quadratic polynomials in the spinor variables, $z^{0,1}$ and their complex conjugate $\bz^{0,1}$. Then their Poisson brackets reproduce the $\so(3,2)$ Lie algebra (see e.g. the appendix of \cite{Dupuis:2011wy}).
To start with, we have already considered the following quadratic polynomials:
\be
\begin{array}{ccl}
J_{z}&=&\f12\,\left[|z^{0}|^{2}-|z^{1}|^{2}\right]
\vspace*{1mm}\,,
\\
K_{+}&=&\f12\,\left[(\bz^{0})^{2}-(z^{1})^{2}\right]
\vspace*{1mm}\,,
\\
K_{-}
&=&\f12\,\left[(z^{0})^{2}-(\bz^{1})^{2}\right]
=\overline{K_{+}}\,,
\end{array}
\ee
which form a $\sl(2,\R)\sim\su(1,1)\sim\so(2,1)$ algebra.
On top of the generator $J_{z}$, one can construct the two remaining $\su(2)\sim\so(3)$ generators:
\be
J_{+}=\bz^{0}z^{1}
\,,\quad
J_{-}=z^{0}\bz^{1}=\overline{J}_{+}\,,
\ee
\be
\{J_{z},J_{\pm}\}=\mp i J_{\pm}
\,,\quad
\{J_{+},J_{-}\}=-2iJ_{z}
\ee
Then, on top of $K_{x}$ and $K_{y}$, we can introduce another boost generator:
\be
K_{z}=-\f12(\bz^{0}\bz^{1}+z^{0}z^{1})
\,,
\ee
so that the six generators, $J_{z},J_{\pm},K_{z},K_{\pm}$ form together a $\sl(2,\C)\sim\so(3,1)$ Lie algebra:
\be
\begin{array}{l}
\{K_{z},K_{\pm}\}=\pm i J_{\pm}
\,,\quad
\{K_{+},K_{-}\}=2i J_{z}
\,,
\vspace*{1mm}\\
\{J_{z},K_{\pm}\}=\mp i K_{\pm}
\,,\quad
\{K_{z},J_{\pm}\}=\mp i K_{\pm}
\,,
\vspace*{1mm}\\
\{J_{+},K_{-}\}=-2iK_{z}
\,,\quad
\{J_{-},K_{+}\}=2iK_{z}
\,,
\vspace*{1mm}\\
\{J_{z},K_{z}\}=\{J_{+},K_{+}\}=\{J_{-},K_{-}\}=0
\,.
\end{array}
\ee
We further introduce another set of boost generators:
\be
\begin{array}{lcl}
L_{z}&=&\f i2\,\big{[}\bz^{0}\bz^{1}-z^{0}z^{1}\big{]}\,,
\vspace*{1mm}
\\
L_{+}&=&-\f {i}2\,\big{[}(\bz^{0})^{2}+(z^{1})^{2}\big{]}\,,
\vspace*{1mm}
\\
L_{-}
&=&\f i2\,\big{[}(z^{0})^{2}+(\bz^{1})^{2}\big{]} =\overline{L}_{+}\,.
\end{array}
\ee
Combining the $K$'s with $L$'s generates the special conformal transformations. In order to close the Lie algebra, we finally have to introduce the dilatation generator for the spinor,
\be
E=\f12\big{[}
\bz^{0}z^{0}+\bz^{1}z^{1}
\big{]}
\,.
\ee
The remaining Poisson brackets are:
\be
\begin{array}{l}
\{L_{z},L_{\pm}\}=\pm i J_{\pm}
\,,\quad
\{L_{+},L_{-}\}=2i J_{z}
\,,
\vspace*{1mm}\\
\{K_{z},L_{z}\}=-E
\,,\quad
\{K_{+},L_{-}\}=\{K_{-},L_{+}\}=-2E
\,,
\vspace*{1mm}
\\
\{K_{z},L_{\pm}\}=\{L_{z},K_{\pm}\}=\{K_{+},L_{+}\}=\{K_{-},L_{-}\}=0
\,,
\vspace*{1mm}
\\
\{J_{z},L_{\pm}\}=\mp i L_{\pm}
\,,\quad
\{L_{z},J_{\pm}\}=\mp i L_{\pm}
\,,
\vspace*{1mm}
\\
\{J_{+},L_{-}\}=-2iL_{z}
\,,\quad
\{J_{-},L_{+}\}=2iL_{z}
\,,
\vspace*{1mm}
\\
\{J_{+},L_{+}\}=\{J_{-},L_{-}\}=0
\,,
\vspace*{1mm}
\\
\{E,J_{a}\}=0
\,,\quad
\{E,K_{a}\}=L_{a}
\,,\quad
\{E,L_{a}\}=-K_{a}
\end{array}
\nn
\ee
From the last line of Poisson brackets, we see that $E$ is actually a  Casimir of the $\su(2)$ Lie algebra. It allows to take the square-root of the quadratic $\su(2)$-Casimir:
\be
J_{z}^{2}+J_{x}^{2}+J_{y}^{2}=J_{z}^{2}+J_{+}J_{-}=E^{2}
\,.
\ee
It also gives the norm of the boost vectors $K$ and $L$:
\be
\label{ortho1}
\vJ^{2}=\vK^{2}=\vL^{2}=E^{2}
\,,
\ee
while the vectors $\vJ$, $\vK$ and $\vL$ are orthonormal:
\be
\vJ\cdot\vK=\vJ\cdot\vL=\vK\cdot\vL=0
\,,
\ee
where the scalar products are computed as $\vJ\cdot\vK=J_{z}K_{z}+\f12(J_{-}K_{+}+J_{+}K_{-})$ in the $z,\pm$ basis.
This means that the 3$\times$3 matrix $E^{-1}(\vJ,\vK,\vL)$ is orthogonal, which leads to equivalent orthonormality conditions between the lines of this matrix, in particular,
\be
\label{ortho2}
\forall a=x,y,z\,,\quad
J_{a}^2+K_{a}^2+L_{a}^2=E^2\,.
\ee
The six orthonormality conditions on the ten $\so(3,2)$ generators reduces the data contained in the $\so(3,2)$ algebra down to the four real components  of the spinor $z\in\C^{2}$. More precisely, an orthonormal $\R^{3}$-basis of  $(\vJ,\vK,\vL)$ with given norm $E$ uniquely\footnotemark{} determines a spinor $z$.
\footnotetext{
The $\su(2)$ generators $J_{a}$ actually give the spinor $z\in\C^{2}$ up to a phase. Then the two other vectors $\vK$ and $\vL$, forming an orthonormal basis with $\vJ$, simply define an angle, which turns out to be the missing phase for uniquely determining the spinor.
}



\section{Cosmological Observables}
\label{sec:so32}

\subsection{Representing the scalar field}

Now we would like to identify those $\so(3,2)$ generators as cosmological observables.
We would like to build on the identification of the $\sl(2,\R)\sim\so(2,1)$ generators $\vcJ=(J_{z},K_{x},K_{y})$ as the CVH observables:
\be
\cH=\f1{2\sigma\ka}(K_{x}-J_{z})
\,,\quad
v=\sigma\ka^3(K_{x}+J_{z})
\,,\quad
\cC=K_{y}
\,.
\nn
\ee
First, we use the $\sl(2,\R)$ Casimir equation \eqref{Csl2R} of the CVH algebra and translate it in terms of the $\so(3,2)$ generators:
\be
-\f{\pip^2}{\ka^2}
=  J^2_z - K_{x}^{2}-K_{y}^{2}
=-L_{z}^{2}
\,.
\ee
It is thus natural to identify the generator $L_{z}$ as the scalar field momentum:
\be
L_{z}=\f{\pip}\ka
\,.
\ee
%
%
Now we wish to identify the variable representing the scalar field $\phi$ itself. It must be canonically conjugate to its momentum, $\{\phi,\pi_{\phi}\}=1$, and also commute with the dilatation generator ${\cC}$, i.e. $\{K_{y},\phi\}=0$.  This actually is not a simple problem. We follow the more pedestrian method of building the whole $\so(3,2)$ algebra of cosmological observables.


%

A crucial point is that we not only want to identify  $\so(3,2)$ generators that satisfy the correct Poisson brackets, but we also want to obtain the same orthonormality conditions as above, thus allowing to identify the cosmological $\so(3,2)$ Lie algebra with the one constructed from the spinor phase space. Indeed, respecting the orthonormality conditions will give the unique mapping between the cosmological variables $(b,v,\phi,\pip)$ and the spinor $z\in\C^{2}$.

First,  we identify the generators $J_{y}$ and $L_{x}$, such that they form a $\so(2,1)\sim\sl(2,\R)$ algebra with $L_{z}$,
\be
\{J_{y},L_{z}\}=L_{x}
\,,\,\,
\{J_{y},L_{x}\}=-L_{z}
\,,\,\,
\{L_{x},L_{z}\}=J_{y}\,.
\ee
They explicitly involve the scalar field $\phi$:
\be
\label{JyLx}
\begin{array}{c}
J_{y}=\f1{2\ka}\Big{[}
(\pip+\ka vb)e^{+\ka \phi}
+
(\pip-\ka vb)e^{-\ka \phi}
\Big{]}
\vspace{1mm}\\
L_{x}=\f1{2\ka}\Big{[}
(\pip+\ka vb)e^{+\ka \phi}
-
(\pip-\ka vb)e^{-\ka \phi}
\Big{]}
\end{array}
\ee
In particular, we check that the Casimir of this $\su(1,1)$ algebra is given by the dilatation generator $\cC=K_{y}$:
\be
J^2_y - L_{x}^{2}-L_{z}^{2}
=-K_{y}^{2}=-v^{2}b^{2}
\,.
\ee
Once we have $J_{y}$ and $J_{z}$, we complete the basic $\su(2)$ algebra with the generator $J_{x}$ computed from the Poisson bracket $\{J_{y},J_{z}\}$:
\beq
J_{x}&=
&
\f{\sigma\ka}{4}\Bigg{[}
-
\left(\f{\pip^{2}}{v}+\ka^{2}v b^{2}- \f v{\sigma^{2}\ka^{4}} +{2\ka\pip b}\right)e^{+\ka \phi}
\nn\\
&&+
\left(\f{\pip^{2}}{v}+\ka^{2}v b^{2}- \f v{\sigma^{2}\ka^{4}} -{2\ka\pip b}\right)e^{-\ka \phi}
\Bigg{]}
\,.
\nn
\eeq
We can check that it is consistent with the other $\su(2)$ Poisson brackets, $\{J_{x},J_{y}\}=J_{z}$ and $\{J_{z},J_{x}\}=J_{y}$.
From here, it is straightforward to complete the whole $\so(3,2)$ algebra, check the Poisson brackets between all the generators and check that they satisfy the same orthonormality relation conditions as wanted.

An important remark is that the expressions of the $\so(3,2)$ generators, as in \eqref{JyLx}, indicates that natural observable for the scalar field is not $\phi$ itself but its exponential $e^{\pm \ka \phi}$, for instance:
\be
\label{simp}
J_{y}\pm L_{x}=
\f1{\ka}
(\pip\pm\ka \cC)e^{\pm\ka \phi}
%
\,.
\ee
This fits with the coupled deparametrized evolution \eqref{vphitraj} of the scalar field and the volume 
and the corresponding  Dirac observables \eqref{DiracVB} (see in appendix \ref{deparametrization} for more details). 

\subsection{Dirac observables from the $\so(3,2)$ algebra}
\label{Obs-so32}

The cosmological Hamiltonian constraint is:
\beq
\label{hh}
\cH
&=&
\f1{2\sigma\ka}(K_{x}-J_{z})
=
\f1{2v}
\left(
\f{\pip}\ka-vb
\right)
\left(
\f{\pip}\ka+vb
\right) \nn\\
&=&
\f1{2\sigma\ka(K_{x}+J_{z})}(L_{z}-K_{y})(L_{z}+K_{y})
\,.
\eeq
This factorization\footnotemark{} reflects that the Hamiltonian constraint  $\cH=0$ is equivalent to  $\pip^{2}=\ka^{2}\cC^{2}$, i.e. $K_{y}^{2}=L_{z}^{2}$.
\footnotetext{
This works because $(K_{x}-J_{z})(K_{x}+J_{z})=(L_{z}-K_{y})(L_{z}+K_{y})$, i.e. $\vK^{2}=J_{z}^{2}+K_{z}^{2}+L_{z}^{2}$ which is equal to $E^{2}$ accordingly to the orthonormality conditions \eqref{ortho1} and \eqref{ortho2} satisfied by the $\so(3,2)$ generators.
}
More precisely, the two branches of solutions, $\pip\simeq \eps\ka vb$ with $\eps=\pm$, corresponding to the expanding and contracting cosmological phases, are given by the constraints $K_{y}=\eps L_{z}$.
There are actually several equivalent expressions of the Hamiltonian constraint, which impose the same condition on the cosmological observables but correspond to different $\SO(3,2)$ flows. For instance, $\cH=0$ is also equivalent to $J_{y}^2=L_{x}^2$. Indeed, as we can see this explicitly from the expressions \eqref{JyLx} of the generators $J_{y}$ and $L_{x}$:
\be
\f\pip\ka=\eps vb
\quad\Leftrightarrow\quad
K_{y}=\eps L_{z}
\quad\Leftrightarrow\quad
J_{y}=\eps L_{x}
\,.
\ee
Although all these conditions are perfectly equivalent at the classical level, the corresponding constraint operators at the quantum level might differ due to different operator ordering. 
A natural question is then which operator seems best suited to implement the dynamics at the quantum level. From this perspective, not only are conditions such as $K_{y}=\eps L_{z}$ and $J_{y}=\eps L_{x}$ straightforward to quantize and solve in $\so(3,2)$ representations, but they further allow to distinguish the two cosmological phases. Nevertheless, the cosmological evolution (in proper time) remains generated by the Hamiltonian constraint $(K_{x}-J_{z})=2\sigma \ka\cH$, which is also straightforward to quantize and solve in $\so(3,2)$ representations. Thus we will keep $K_{x}-J_{z}=0$ as the legitimate Hamiltonian constraint at the quantum level, with the other conditions, $K_{y}-\eps L_{z}=0$ and $J_{y}-\eps L_{x}=0$, nevertheless remaining relevant to determine the trajectories and deparametrize the theory at the quantum level.

\medskip

Let us now look for Dirac observables and, in particular identify how to recover the ``evolving'' Dirac observables $\cV_{\eps}$ and $\cB_{\eps}$.
The $\so(3,2)$ algebra allows for a more systematic search for Dirac observables. Checking among Lie algebra elements,  we  identify a pair of commuting strong Dirac observables:
\be
\begin{array}{c}
\{\cH,K_{z}+J_{x}\}
=
\{\cH,E+L_{y}\}
=
0
\,,
\vspace{1mm}\\
\{K_{z}+J_{x},E+L_{y}\}=0
\,.
\end{array}
\ee
It is fairly easy to compute\footnotemark{} these two observables explicitly in terms of $b,v,\phi,\pip$:
\be
\label{Diracobs}
\begin{array}{lcl}
E+L_{y}
&=&
\f{\sigma \ka}{2v}\Big{[}
e^{\ka\phi}\big{(}\pip+\ka vb\big{)}^2
+
e^{-\ka\phi}\big{(}\pip-\ka vb\big{)}^2
\Big{]}\,,
\vspace{2mm}\\
K_{z}+J_{x}
&=&
\f{\sigma \ka}{2v}\Big{[}
-e^{\ka\phi}\big{(}\pip+\ka vb\big{)}^2
+
e^{-\ka\phi}\big{(}\pip-\ka vb\big{)}^2
\Big{]}\,.
\end{array}
\nn
\ee
\footnotetext{
The simplest way is to compute $K_{z}=\{J_{x},\cC\}$, $L_{y}=\f1\ka\{\pip,J_{x}\}$ and finally $E=\{L_{y},\cC\}$.
}
From these expressions, one can directly check that they define strong Dirac observables, whose bracket with the Hamiltonian constraint $\cH$ exactly vanish. Then, if we assume to be on the branch $\pip\simeq \eps\ka\cC$ of solutions to the Hamiltonian constraint, then these observables reduce to the volume observable $\cV_{\eps}$:
\be
\pip\simeq \eps\ka\cC
\,\Rightarrow\,
(E+L_{y})\simeq{2\sigma\ka\pip^{2}}\,\f1{ve^{-\eps\ka\phi}}\simeq-\eps(K_{z}+J_{x})
\,.
\nn
\ee
From this point of view,  the observable $(E+L_{y})$ combines the two weak observables $\cV_{\pm}$ corresponding to the two branches into a single strong Dirac observable. Moreover, since it is a $\so(3,2)$ Lie algebra element, it will admit a straightforward ambiguity-free quantization, when we will raise the spinor components to creation and annihilation operators.

It is natural to introduce the sum and difference of these two (commuting) Dirac observables:
\beq
\label{classicalApm}
A_{\pm}
&=&
(E+L_{y})\pm(K_{z}+J_{x})
\\
&=&
\f{\sigma\ka}{v}e^{\mp \ka \phi}\,\big{(}\pip\mp\ka vb\big{)}^2\,.\nn
\eeq
Depending on the phase, contracting or expanding, one of the two observables $A_{+}$ or $A_{-}$ vanishes while the other indicates the value of the volume $v$ at the initial condition $\phi=0$.

It is remarkable that they form a very simple algebra of Dirac observables together with the scalar field momentum $L_{z}=\pip/\ka$:
\be
\{L_{z},A_{\pm}\}=\pm A_{\pm}
\,,\quad
\{A_{+},A_{-}\}=0
\,.
\ee
On the one hand, the $\sl(2,R)$ Casimir $L_{z}=\cE=\pip/\ka$ acts as the generator of dilatation on the observables $A_{\pm}$,
\be
e^{\{\eta L_{z}, \cdot\}}A_{\pm}= e^{\pm \eta}A_{\pm}
\ee
On the other hand, it means that the DIrac observables $A_{\pm}$ generates shifts of the $\sl(2,R)$ Casimir, i.e. of the scalar field momentum $\pip$,
\be
e^{\{\zeta A_{\pm}, \cdot\}} L_{z}
=
L_{z}\mp\zeta A_{\pm}
\,.
\ee
At the quantum level, the cosmological systems will be quantized as a $\sl(2,R)$ representation, with each irreducible representation corresponding to a given value of the scalar field momentum $\pip$. The operators $\hA_{\pm}$  will then generate shifts in the $\sl(2,R)$ Casimir, allowing to transition between irreducible $\sl(2,\R)$-representations.
This reflects the fact that the  observables $A_{\pm}$ depends on the scalar field $\phi$. Not only do such operators  allow to explore the space of FRW trajectories for a free massless scalar field by changing the value of $\pip$,  but they also open the door to adding a scalar field potential (explicitly depending on $\phi$) and representing it at the quantum level in terms of $\so(3,2)$ operators.

\medskip

To conclude the description of the $\so(3,2)$ operators at the classical level, it is interesting to look at the action of the Dirac observables $A_{\pm}$ not only on the $\sl(2,\R)$ Casimir $L_{z}=\pip/\ka$ but also directly on the CVH observables.
A quick application of the $\so(3,2)$ Poisson brackets gives
%
%
\be
\begin{array}{lcl}
\{A_{\pm}, K_{y}\}&=&A_{\pm}
\,,\\
\{A_{\pm}, K_{x}-J_{z}\}&=&0
\,,\\
\{A_{\pm}, K_{x}+J_{z}\}&=&2(L_{x}\mp J_{y})
\,,\\
\{A_{\pm},L_{x}\mp J_{y}\}&=&0
\,,
\end{array}
\ee
%
%
where we recall the mapping betwen $\sl(2,\R)$ generators and the cosmological observables with $\cC=K_{y}$, $\cH=(K_{x}-J_{z})/2\sigma\ka$ and $v=\sigma\ka^3(K_{x}+J_{z})$. One can then complete this algebra with the Poisson bracket $\{A_{\pm},(L_{x}\pm J_{y})\}=-2(K_{x}-J_{z})$.
Looking at those relations more closely, the observables $A_{\pm}$ obviously commute with the Hamiltonian constraint -they are Dirac observables- as already derived earlier, and they simply induce shifts in the dilatation generator $\cC$ and in the volume $v$:
\be
\begin{array}{lcl}
e^{\{\zeta A_{\pm}, \cdot\}} \cC
&=&
\cC+\zeta A_{\pm}
\,,\\
e^{\{\zeta A_{\pm}, \cdot\}} v
&=&
v+2\zeta\sigma\ka^3 (L_{x}\mp J_{y})
\,.
\end{array}
\ee
Let us look at those variations on-shell when applied to physical cosmological trajectories. For instance, considering the expanding phase, thus with $\pip=+\ka\cC$ or equivalently $K_{y}=L_{z}$, one also has $A_{+}=0$ and $J_{y}=L_{x}$. 
First, the value of $A_{-}$ gives the deparametrized value of the volume,
\be
\label{aaaa}
A_{-}=
\f{\sigma\ka}{v}e^{\mp \ka \phi}\,\big{(}\pip+\ka vb\big{)}^2
\underset{\pip=\ka\cC}{\simeq}
\f{4\sigma\ka\pip^{2}e^{\ka \phi}}{v}
\,.
\ee
Second, acting with 
$e^{\{\zeta A_{-}, \cdot\}}$ on the CVH observables gives the finite variations\footnotemark:
\be
\begin{array}{lcl}
\Delta^{-}_{\zeta} \cH &=&0
\,,\\
\Delta^{-}_{\zeta} \pip&=& \zeta \ka A_{-}\,,\\
\Delta^{-}_{\zeta} \cC&=&\zeta A_{-}\,,\\
\Delta^{-}_{\zeta} v &=& 2\zeta\sigma\ka^3 (L_{x}+ J_{y})
\sim 4\zeta\sigma\ka^2\pip e^{+\ka\phi}
\,.
\end{array}
\ee
\footnotetext{
We define the finite variation for an observable $\cO$ by $\Delta^{-}_{\zeta} \cO\equiv e^{\{\zeta A_{-}, \cdot\}}\cO- \cO$, by contrast with the infinitesimal variation $\delta^{-} \cO\equiv \{ A_{-}, \cO\}$, which is simply the leading order coefficient of the Taylor expansion of the finite variation.
}
This action preserves the  condition characterizing the physical expanding phase, $\pip=+\ka\cC$, and it shifts the volume while keeping a constant value of the Dirac observable $A_{-}$.
More precisely, if we start with a physical expanding trajectory at a given value of $A_{-}$, satisfying by definition the relation  $\pip=+\ka\cC$ and thus the Hamiltonian constraint $\cH=0$, the action of $e^{\{\zeta A_{-}, \cdot\}}$ will generate another physical expanding trajectory with a different value of the scalar field momentum $\pi_{\phi}$, but the same value for  $A_{-}$. If the initial trajectory is given by the deparametrized equation $ve^{-\ka\phi}=4\sigma\ka\pi_{\phi}^{2}/A_{-}$ for the initial value the scalar field momentum $\pi_{\phi}$, then the final trajectory is given by $ve^{-\ka\phi}=4\sigma\ka\tilde{\pi}_{\phi}^{2}/A_{-}$ for the final value the scalar field momentum $\tilde{\pi}_{\phi}$.
The volume $v$ and the scalar field $e^{+\ka\phi}$ gets shifted accordingly from the initial trajectory to the final trajectory. 

Third, the flow induced by $A_{+}$ is trivial,
\be
\Delta^{+}_{\zeta} \cH 
=
\Delta^{+}_{\zeta} \pip
=
\Delta^{+}_{\zeta} \cC
=
\Delta^{+}_{\zeta} v 
=
0
\,.
\ee
We get the same structure for the contracting phase by exchanging the role of $A_{-}$ and $A_{+}$.
So the flow generated by the Dirac observables $A_{\pm}$ on physical cosmological trajectories defines a change of trajectories given by a shift in the scalar field momentum $\pip$, while adjusting the volume.

\medskip

Through the exploration of the $\so(3,2)$ algebra of observables for the FRW cosmology of a massless free scalar field, we have seen how it enriches the $\sl(2,\R)$ structure of the CVH algebra previously introduced and discussed in \cite{BenAchour:2017qpb,BenAchour:2018jwq, BenAchour:2019ywl,BenAchour:2019ufa}. Accounting for the scalar field $\phi$, and not treating its momentum $\pip$ as a mere coupling constant for the dynamics of the geometry, allows one to include in this $\so(3,2)$ algebra Dirac observables encoding the coupled evolution of the scalar field and volume (thus representing the deparametrized volume) and define Hamiltonian flows not only along the cosmological trajectories but also transversal to them.
Although introducing such flows changing the trajectory may appear as a complicated sophistication for a massless free scalar field, it will become necessary when including a mass or non-trivial potential as it shall lead to physical trajectories deviating from the present massless free cosmological trajectories.

All these classical algebraic structures and Hamiltonian flows, with the $\sl(2,\R)$ symmetry, $\so(3,2)$ brackets and the Dirac observables, will be straightforwardly preserved by the quantization.


\subsection{Adding a self-interaction potential}
\label{phipotential}

Having observables depending on the scalar field  at our disposal, we now suggest how to use this new feature to investigate relevant inflationary potentials for $\phi$ at the quantum level.
%
Adding a potential to the original action, the scalar Hamiltonian constraint then becomes:
\be
\cH = \frac{\pi^2_{\phi}}{2v} - \frac{\kappa^2}{2} v b^2 + \frac{V(\phi)}{3\kappa^2} v = \f1{2\sigma\ka}(K_{x}-J_{z}) + \frac{V(\phi)}{3\kappa^2} v
\,.
\ee
The natural question is then if the potential term in $vV(\phi)$ can be easily written in terms of the $\so(3,2)$ observables.
As it turns out, most of the relevant self-interacting potentials used in single field inflationary models can be written as polynomial expressions of the exponential function $e^{-\kappa \phi}$ (see \cite{Martin:2013nzq} for an exhaustive review of the most relevant inflationary potentials). Interesting examples of such potentials are given by
\be
\left|\begin{array}{lcl}
V_{\text{Starobinsky}}(\phi)  &\propto& \left(1 - e^{- \lambda\kappa \phi} \right)^2 \;, \vspace*{2mm}\\
V_{\text{power law}}(\phi)  &\propto &\; e^{-\lambda \kappa \phi} \;, \vspace*{2mm}\\
V_{\text{ESUSY}}(\phi)  &\propto& \left( 1 - e^{-\lambda \kappa \phi} \right)\;,
\end{array}\right.
\ee
%
which correspond respectively to the well-known Starobinsky (or Higgs) inflation model favored by the current data, to power law inflation, and finally to the exponential SUSY inflation model.
Depending on the model, $\lambda$ takes an a priori specific value of order 1 (as for the Starobinsky potential) or an arbitrary real value to be determined or fine-tuned a posteriori after comparing the predictions with the experimental data.

Here, we would like to point out that the building block function $e^{\pm \kappa \phi}$ can be naturally written as a suitable combination of $\so(3,2)$ generators. There are several ways to recast this function, using for instance \eqref{simp} or \eqref{Diracobs}. The simplest expression is obtained by the following linear combinations:
\be
J_{x}-K_{z}=
\f{v}{\sigma \ka^{3}}\sinh\ka\phi
\,,\qquad
E-L_{y}=
\f{v}{\sigma \ka^{3}}\cosh\ka\phi
\,,
\ee
from which we easily extract $ve^{\pm\ka\phi}$ as linear combinations of $\so(3,2)$ generators. Thus although the self-interaction term $vV(\phi)$ might not directly be in the $\so(3,2)$ Lie algebra of observables, it can certainly be written as a polynomial or rational function of the $\so(3,2)$ generators\footnotemark{} (or more generally as a power of the $\so(3,2)$ generators in the case of the power law model for instance) when the parameter $\lambda$ is an integer. In the general case when $\lambda$ is an arbitrary real number, we face the usual issue and quantization ambiguities arising when dealing with  a non-rational or non-analytic potential.
\footnotetext{
Let us point out that there are hints of a more general conformal structure with a Virasoro algebra extending the $\sl(2,\R)$ structure \cite{Lidsey:2012bb,Lidsey:2018byv}. Virasoro group transformations act non-trivially on the potential, or in other words it looks possible to write the scalar field potential in terms of Virasoro generators. This seems in the same line of thought as the mapping of cosmology onto conformal mechanics. It would be interesting to investigate this further, as well as the connection of that formalism with the $\so(3,2)$ structure studied here.
}

Nevertheless, we would like to underline that the main point of introducing the  $\so(3,2)$  observables was to remedy the fact that the $\sl(2,\R)$ observables did not involve at all the scalar field $\phi$ and thus that they did not allow to introduce a potential. Now, the $\so(3,2)$  observables allow to introduce and represent a potential even though its expression might not be simple.

So if one tackles  the canonical quantization of this self-interacting system directly for arbitrary real $\lambda$, the $\so(3,2)$ structure may not be of special help. However, in the context of effective approaches such as the quantum phase space method, the $\so(3,2)$ structure discussed here might provide a simplifying ingredient to compute the higher-moments of the wave function, and discuss the back-reaction of fluctuations and correlation on the inflationary quantum dynamics. See \cite{Bojowald:2010qm, Brizuela:2015fwa, Brizuela:2014cfa, Alonso-Serrano:2020szo, Brizuela:2019elx} for recent applications of this powerful framework.


\section{Quantum Cosmology from $\SL(2,\R)$ Representations} 
\label{su11HO}

We move on to the quantum theory.
The standard method to quantize FRW cosmology is to choose a polarization for the gravitational sector and consider either wave-functions $\Psi(v,\phi)$ or $\Psi(b,\phi)$. One would then translate the Hamiltonian constraint and other observables into differential operators by applying the standard canonical quantization prescription with $\hat{v}$ and $\hat{b}$ quantized as multiplicative or derivative operators. One could also apply an alternative quantization procedure in an attempt to further regularize the behavior of the theory close to the cosmological singularity, such as the polymer quantization used in loop quantum cosmology \cite{Ashtekar:2011ni, Bojowald:2008zzb, Agullo:2016tjh}. Here we keep the standard canonical quantization, but apply it to the spinorial phase space.

Indeed, having parametrized the cosmological phase space with complex variables and mapped it onto the phase space $\C^{2}$, we simply quantize the spinor $z\in\C^{2}$ as a pair of harmonic oscillator and promote  the spinor components $z^{0,1}$ and $\bz^{0,1}$ to annihilation and creation operators:
\be
\left|\begin{array}{lcl}
z^{0}&\rightarrow &a\\
z^{1}&\rightarrow &b\\
\bz^{0}&\rightarrow &a^{\dagger}\\
\bz^{1}&\rightarrow &b^{\dagger}
\end{array}\right.
\qquad\textrm{with}\quad
\begin{array}{l}
[a,a^{\dagger}]=[b,b^{\dagger}]=1\,,
\vspace{1mm}\\
{[}a,b{]}=0\,.
\end{array}
\ee
Considering the quadratic polynomials in those harmonic oscillators turns the classical $\so(3,2)$ generators into quantum operators. We then translate them in cosmological observables.

Such a simple scheme has two immediate advantages. First, we can use the normal operator ordering for harmonic oscillators, thereby fixing all quantization ambiguities. Second, it automatically preserves the $\so(3,2)$ algebra. In particular, it preserves the $\sl(2,\R)$ sub-algebra, thus conserving at the quantum level the conformal symmetry of FRW cosmology under Mobius transformation of the proper time, as given classically \eqref{Mobius}. A side-product of realizing the $\so(3,2)$ algebra without anomaly at the quantum level implies that we preserve the Dirac observables identified classically, $L_{z}$ and $A_{\pm}$, the Lie algebra that they form and Hamiltonian flows that they generate.


\subsection{Quantum cosmology from harmonic oscillators}

Let us start by revisiting the quantization of the CVH observables.
The canonical quantization of the spinor (components) leads to the quantization of the $\sl(2,\R)$ generators as:
\be
\begin{array}{ll}
\hJ_{z}=\f12(a^{\dagger}a-b^{\dagger}b)
\,,\,&
\hcE=\hL_{z}=\f  i2(a^{\dagger}b^{\dagger}-ab)
\,,
\vspace{1mm}\\
\hK_{+}=\f12\left((a^{\dagger})^{2}-b^{2}\right)
\,,\,&
\hK_{-}
=
\f12\left(a^{2}-(b^{\dagger})^{2}\right)
\,.
\end{array}
\ee
where $\hJ_{z}$ measures the energy difference between the two harmonic oscillators, while $\hK_{\pm}$ and the $\hL_{z}$ are squeezing operators. They satisfy the  $\sl(2,\R)$ commutators, without corrections:
\be
[\hJ_{z},\hK_{\pm}]=\pm \hK_{\pm}
\,,\quad
[\hK_{+},\hK_{-}]=-2\hJ_{z}
\,,
\ee
while the quadratic Casimir acquires an expected $\f14$ constant correction due to the non-commutativity of the generators:
\be
\label{Cvalue}
\hC=\hJ_{z}^{2}-\f12(\hK_{-}\hK_{+}+\hK_{+}\hK_{-})
=-\left(
\hcE^{2}+\f14
\right)\,.
\ee
These four operators allow to represent at the quantum level the volume $v$, the dilatation generator $\cC$, the Hamiltonian constraint $\cH$ and the scalar field momentum $\pip$ as
\be
\begin{array}{ll}
\hv=\sigma\ka^{3}(\hK_{x}+\hJ_{z})
\,,\quad&
\hcC=\hK_{y}
\,,
\vspace{1mm}\\
\hcH=\f1{2\sigma\ka}(\hK_{x}-\hJ_{z})
\,,\quad&
\hpi=\ka\hcE=\ka\hL_{z}
\,.
\end{array}
\ee
We  extend this canonical quantization scheme to the whole $\so(3,2)$ algebra and raise all the $\so(3,2)$ generators to quantum operators $\hJ_{a}$, $\hK_{a}$, $\hL_{a}$ and $\hE$, thereby also constructing operators involving the scalar field $\phi$. We don't write them explicitly here, although it is straightforward to check that all the commutators of the  $\so(3,2)$ algebra are preserved.
%
In particular, the operators $\hL_{z}=\hcE$ (the scalar field momentum) and $\hA_{\pm}$ (constructed from $\hE+\hL_{y}$ and $\hK_{z}+\hJ_{x}$) still provide strong Dirac observables, commuting with the Hamiltonian constraint operator $\hcH$.

\medskip

The Hilbert space consists in the tensor product of two copies of the Hilbert space of an harmonic oscillator, with states labeled by the energy levels of the two oscillators $a$ and $b$:
\be
\hil=\hil_{HO}^{\otimes 2}=\bigoplus_{n_{a},n_{b}\in\N}\C\,|n_{a},n_{b}\ra\,.
\ee
This Hilbert space carries a $\so(3,2)$-representation, which can actually be decomposed as a ladder of irreducible representations of its $\sl(2,\R)$ subalgebra.

Upon diagonalizing the $\sl(2,\R)$ Casimir operator $\hL_{z}=\hcE$ and the $\mathfrak{u}(1)$ generator $\hJ_{z}$, it can be decomposed in irreducible $\SL(2,\R)$-representations  with states labeled by the eigenvalues of $\hL_{z}$ and $\hJ_{z}$. Irreducible unitary $\SL(2,\R)$-representations are reviewed in appendix \ref{su11irrep}. From the expression \eqref{Cvalue} of the quadratic Casimir, we see that we reach only representations from the principal continues series, also referred to as space-like representations. They are labeled by the eigenvalue $s\in\R$ of the operator $\hcE=\hL_{z}$ and have a strictly negative quadratic Casimir, $\hC=-(s^{2}+\f14)$:
\be
\hil=\int_{\R} \rd s\,\bigoplus_{m\in\Z}\C\,|s,m\ra\,.
\ee
Since the scalar field momentum operator $\hpi$ is simply $\ka$ times $\hL_{z}$, it is diagonal in this continuous basis:
\be
\hpi\,|s,m\ra=\,\ka\,s\,|s,m\ra
\,.
\ee
Note that the operator $\hJ_{z}$ is not directly the volume $\hv$ and involves a linear combination of $\hv$ with the Hamiltonian constraint $\hcH$. In particular, their action on physical states, annihilated by $\hcH$, will be the same but their action and commutation on arbitrary states generally do not match.

\medskip

Going beyond $\sl(2,\R)$ representations, we expect an operator representing the scalar field $\phi$ to create shifts in the  $\sl(2,\R)$ representation label $s$ and  transitions between $\SL(2,\R)$-representations. As we have seen in the description of the classical $\so(3,2)$ algebra, we don't have direct access to the scalar field. Nevertheless all the $\so(3,2)$-generators, apart from the $\sl(2,\R)$-generators, involve the scalar $\phi$ and allow to define $\phi$-dependent  operators acting on the Hilbert space of cosmological quantum states $\hil$.


In the present setting exploring the simplest case of the cosmology of a massless free scalar field, one can legitimately work in a subspace at fixed $s=\pip/\ka$ in a single irreducible  $\SL(2,\R)$-representation. However, as soon as one adds a mass or turns on a potential (e.g. to study inflation scenarios), the dynamics will necessarily involve hopping from one irreducible  $\SL(2,\R)$-representation to another, and physical states will be spread out on $\hil$ and won't live in a single subspace at fixed $s$.


\subsection{Dirac observable operators}

Once the quantization of the phase space has been achieved and the Hilbert space of quantum states for the coupled cosmological system geometry plus scalar matter has been defined, the next step is to determine physical states solving the Hamiltonian constraint operator and to investigate the eigenstates and flow of the Dirac observables.

Let us start with determining solutions of the Hamiltonian constraint, i.e. eigenstates of $\hcH$ with vanishing eigenvalues, thus satisfying $\hK_{x}|\Psi\ra=\hJ_{z}|\Psi\ra$. In \cite{BenAchour:2019ywl}, this equation is solved on a given $\sl(2,\R)$-representation at fixed label $s=\pip/\ka$. Decomposing the state on the basis diagonalizing $\hJ_{z}$, 
\be
|\Psi^{(s)}\ra=\sum_{m\in\Z}\Psi^{(s)}_{m}\,|s,m\ra
\,,
\nn
\ee
the equation $\hK_{x}|\Psi\ra=\hJ_{z}|\Psi\ra$ turns into a second order recursion relation on the coefficients $\Psi^{(s)}_{m}$, which can be solved using the method by Laplace (e.g. \cite{AIHPA_1970__13_1_27_0}), and leads to two independent solutions respectively with support on even $m\in2\Z$ and on odd $m\in2\Z$+1 . For large $m$, both solutions produce scale invariant oscillations, $\Psi^{(s)}_{m}\simeq \exp[is\ln|m|]$. These are the physical states for fixed scalar field momentum $\pip=\ka s$. This amounts to simultaneously diagonalizing the Hamiltonian constraint and the Dirac observables $\hL_{z}$. Indeed the value of $\hL_{z}=\hat{\pi}_{\phi}/\ka$ determines a unique cosmological trajectory. 

Instead of $\hL_{z}$, we could focus on the other Dirac observables that we constructed,
\be
\hA_{\pm}=(\hE+\hL_{y})\pm(\hK_{z}+\hJ_{x})
\,,
\ee 
Classically, they give the value of the deparametrized volume $ve^{\pm \ka \phi}$ as shown in \eqref{classicalApm}. One can not simultaneously diagonalize the operators $\hL_{z}$ and $\hA_{\pm}$ since their commutator does not vanish:
\be
[\hL_{z},\hA_{\pm}]=\pm i \hA_{\pm}
\,,\quad
[\hA_{+},\hA_{-}]=0
\,.
\ee
The operators $A_{\pm}$ clearly do not act on a given  $\sl(2,\R)$-representation at fixed label $s$, but hop between irreducible representations. Simultaneously diagonalizing the Hamiltonian constraint and the Dirac observables $\hA_{\pm}$ will thus lead to physical states $|\Phi^{(a_{\pm})}\ra$ which are (coherent) superpositions of the states $|\Psi^{(s)}\ra$. At the classical level, this corresponds to the fact that the values of $A_{\pm}$ determine a unique cosmological trajectory. 
From this perspective, using the Dirac observables $\hL_{z}$ or $A_{\pm}$ corresponds to studying different statistical cosmological ensembles, depending whether trajectories are identified by the scalar field momentum $\pip$ or the value of the volume at a given point of the evolution (at $\phi=0$).

\medskip

Revisiting this discussion from the point of view of the pair of harmonic oscillators used to build the $\so(3,2)$ representation, the Hamiltonian constraint operators reads:
\beq
\hcH&\propto& 2(\hK_{x}-\hJ_{z})\\
&=& \f12\big{[}a^{2}+(a^{\dagger})^{2}-b^{2}-(b^{\dagger})^{2}\big{]}-a^{\dagger}a+b^{\dagger}b\,.\nn
\eeq
If we decompose states on the harmonic oscillator basis,
\be
|\Psi\ra=\sum_{n_{a},n_{b}\in\N}\Psi_{n_{a},n_{b}}\,|n_{a},n_{b}\ra\,,
\nn
\ee
the vanishing eigenvalue problem $\hcH|\Psi\ra=0$ turns into a double 2nd-order equation determining coupled squeezed states for the two oscillators:
\beq
&&\sqrt{(n_{a}+1)(n_{a}+2)}\Psi_{n_{a}+2,n_{b}}
+\sqrt{n_{a}(n_{a}-1)}\Psi_{n_{a}-2,n_{b}}\nn\\
&&-\sqrt{(n_{b}+1)(n_{b}+2)}\Psi_{n_{a},n_{b}+2}
-\sqrt{n_{b}(n_{b}-1)}\Psi_{n_{a},n_{b}-2}\nn\\
&&+2(n_{a}^{2}-n_{b}^{2})\Psi_{n_{a},n_{b}}
=0
\,.
\eeq
The infinite number of independent initial conditions for this recursion relation, depending on the values for $\Psi_{0,n_{b}}$, $\Psi_{1,n_{b}}$, $\Psi_{n_{a},0}$ and $\Psi_{n_{a},1}$, lead to an infinity of physical cosmological states, which can then be distinguished by the values of the Dirac observable operators $\hL_{z}$ or $\hA_{\pm}$.

Instead of solving this equation, one could alternatively use the coherent state basis (or a squeezed state basis) for the harmonic oscillators. The $\so(3,2)$ operators are squeezing operators. One can then study the cosmological evolution and flows of the Dirac observables on those states $|z_{a},z_{b}\ra$, recovering the classical spinorial formalism introduced in this work plus quantum corrections.


\newpage

\section*{Conclusion}

This technical paper extends the $\sl(2,\R)$ formalism for FRW cosmology introduced in \cite{BenAchour:2017qpb,BenAchour:2018jwq,BenAchour:2019ywl} and proven to reflect the conformal symmetry for the homogeneous sector of gravity coupled to a massless free scalar field \cite{BenAchour:2019ufa}. Indeed, the $\sl(2,\R)$ Lie algebra describes the observables for the gravitational sector of the theory at fixed scalar field momentum.  Here we extended it to a $\so(3,2)$ algebra of observables coupling gravity to the scalar field. Instead of focusing on the AdS${}_{2}$ phase space for the geometry, this allows one to represent the whole phase space including the scalar field. Having observables explicitly depending on the scalar field allows to represent a full set of Dirac observables, including the deparametrized volume (which encodes the evolution of the volume with respect to the scalar field used as an internal clock), and opens the door to adding a scalar field mass or potential to the action.

The outlook of this work is mostly towards applying this new tool to further explore the physics of FLRW quantum cosmology.
A first question concerns possible new hidden symmetry of FRW cosmology. Indeed, the $\sl(2,\R)$ generators, identified as the volume, the extrinsic curvature and Hamiltonian of the gravity plus matter system, have already been understood as the Noether charges of the conformal symmetry of the action under Mobius transformations of the proper time \cite{BenAchour:2019ufa}. Since the $\so(3,2)$ generators contain Dirac observables (involving the scalar field $\phi$), this might reflect a potentially larger conformal invariance of FRW cosmology. Therefore, is there a symmetry behind the $\so(3,2)$ algebra of observables discussed here ?
If yes, this would be similar in spirit to the  $\so(4,2)$ conformal symmetry of the hydrogen atom, allowing for a full exact solution at the quantum level.

A second topic of investigation would be the application of the $\so(3,2)$ toolbox and  spinorial formalism to the coarse-graining of the coupled geometry+matter along the lines of \cite{Bodendorfer:2019wik, Bodendorfer:2018csn}, which proposed using the $\sl(2,\R)$ structure as a guide to coarse-grain the quantum states of the geometry.

Third, perhaps the most important outlook is the application to the study of non-trivial potentials, for instance in the context of inflation. In particular, it remains to investigate whether the $\so(3,2)$ symmetry is a generic feature of those models and if it allows for a non-ambiguous quantization of inflationary backgrounds using conformal bootstrap techniques, as done in the massless free case \cite{BenAchour:2019ufa}, or at least for a perturbative scheme to study the behavior around the initial cosmological singularity. Nevertheless, we point that effective approaches to quantum cosmology, such as the quantum phase space formalism discussed for examples in \cite{Bojowald:2010qm, Brizuela:2015fwa, Brizuela:2014cfa, Alonso-Serrano:2020szo, Brizuela:2019elx}, combined to the new $\so(3,2)$ structure presented in this work, shall provide a powerful framework to investigate the quantum dynamics of inflationary backgrounds and the quantum back-reaction of the higher moments of the wave function. We leave these open directions for future works.

\section*{Acknowledgements}
The work of J.BA was supported by Japan Society for the Promotion of Science Grants-in-Aid for Scientific Research No. 17H02890.

\newpage


\appendix

\section{Deparametrization and Dirac Observables} 
\label{deparametrization}

\subsection{Deparametrizing and the Scalar Field Time}

Homogeneous and isotropic FLRW cosmology with a scalar field consists in two canonical pair of variables, the volume with its conjugate variable $\{b,v\}=1$ describing the geometry sector and the scalar field with its conjugate momentum $\{\phi,\pi_{\phi}\}=1$ describing the matter sector, provided with a Hamiltonian constraint $\cH$. This Hamiltonian constraint generates the time evolution (or in more technical terms, diffeomorphisms in the time direction) and implies the theory's invariance under time re-parametrization. Since the coordinate time $t$ (for lapse $N=1$) is not physical, the physical content of the theory is more conveniently described by an evolution with respect to a internal clock. The typical choice is to use the scalar field $\phi$ as a clock and to describe the evolution of the geometry $(b,v)$ in terms of $\phi$.

The effective Hamiltonian for this evolution with respect to the scalar field clock is its conjugate momentum $\pi_{\phi}$ once the Hamiltonian constraint is solved\footnotemark:
\be
\cH = \frac{\pi^2_{\phi}}{2 v} - \frac{\ka^2v b^2}{2} 
=\f1{2 v}(\pip-\ka\cC)(\pip+\ka\cC)
=0
\ee
\footnotetext{The evolution with respect of the coordinate time $t$ is given by the Hamiltonian equations,$\pp_{t}\phi=\{\phi,\cH\}$ for the scalar field and $\pp_{t}\cO=\{\cO,\cH\}$ for an observable $\cO(v,b)$ probing exclusively the geometry sector.
%
The deparametrized evolution is given by the ratio
\be
\f{\dd \cO}{\dd\phi}=\f{\pp_{t}\cO}{\pp_{t}\phi}=\f{\{\cO,\cH\}}{\{\phi,\cH\}}
\,.
\nn
\ee
Assuming that the Hamiltonian constraint is quadratic in the scalar field momentum and does not depend on the scalar field itself, i.e. of the form $\cH=f(\pi_{\phi}-a)(\pi_{\phi}-b)$ where $f,a,b$ are phase space functions depending only on the geometry variables $v$ and $b$, we easily compute the deparametrized evolution, assuming e.g. that we are on the branch $\pi_{\phi}=a$:
\be
\f{\dd \cO}{\dd\phi}=\f{f(a-b)\{a,\cO\}}{f(a-b)}=\{a,\cO\}\,,
\nn
\ee
which means that the effective Hamiltonian for the deparametrized evolution on the branch $\pi_{\phi}=a$ is simply $\Heff=a$.
}
Therefore we have two branches for the evolution with respect to $\phi$, with the effective Hamiltonian given by the dilatation generator $\cC$ up to a numerical factor and up to a sign:
\be
\Heff=\pi_{\phi}
=\pm\ka\,\cC\,.
\ee
This gives the evolution of the geometry sector (with respect to the matter sector) as the flow generated by the dilatation generator. Choosing the positive branch, $\Heff=\ka\,\cC$, we get:
\be
\label{deparamtraj}
\left|\begin{array}{lclcl}
v(\phi)&=&e^{\phi\{\Heff,\cdot\}}v&=&v^{(0)}e^{\ka\phi}
\\
b(\phi)&=&e^{\phi\{\Heff,\cdot\}}b&=&b^{(0)}e^{-\ka\phi}
\end{array}\right.
\ee
where $(b^{(0)},v^{(0)})$ are the initial values at $\phi=0$. First, we recover the cosmological trajectories \eqref{vphitraj} computed above. Then, we check that $\cC$ is an constant of motion, $v(\phi)b(\phi)=b^{(0)}v^{(0)}$, since it is actually the Hamiltonian evolving the system. We could also recover those same trajectories from the $\SU(1,1)$ flow generated by the dilatation $\cC=k_{y}$ on the pure gravity sector\footnotemark.
\footnotetext{
As a consistency check, we can exponentiate the action of the effective Hamiltonian  $\Heff=\ka\,\cC$ in terms of $\SU(1,1)$ group elements. Here, we act on the pure gravity sector and we work with the $\su(1,1)$ generators $\vj=(j_{z},k_{x},k_{y})$:
\be
m=\mat{cc}{j_{z}&k_{-}\\k_{+}& j_{z}}\,,\quad
e^{\eta\{\cC,\cdot\}}m=G_{\eta}mG_{\eta}^\dagger
\,,
\nn
\ee
\be
G_{\eta}=e^{i\f\eta2\tau_{y}}=\mat{cc}{\cosh\f\eta2 & \sinh\f\eta2 \\ \sinh\f\eta2 & \cosh\f\eta2}\,,
\nn
\ee
where $\eta=\ka\,\phi$ is the scalar field clock. Computing the resulting evolution for the $\su(1,1)$ generators allows to recover the expected trajectories for the volume $v$ and the Hamiltonian constraint:
\be
\left|\begin{array}{lcl}
j_{z}(\eta)&=&j_{z}^{(0)}\cosh\eta+k_{x}^{(0)}\sinh\eta\\
k_{x}(\eta)&=&k_{x}^{(0)}\cosh\eta+j_{z}^{(0)}\sinh\eta\\
k_{y}(\eta)&=&k_{y}^{(0)}
\end{array}\right.
\Rightarrow
\left|\begin{array}{lcl}
v(\eta)&=&v^{(0)}e^{\eta}\\
\cH_{g}(\eta)&=&\cH_{g}^{(0)}e^{-\eta}\\
\cC(\eta)&=&\cC^{(0)}
\end{array}\right.
\nn
\ee
The dilatation generator stays constant, the Hamiltonian constraint always vanishes if its initial value vanishes and the volume follows an exponential expansion.
}

At the end of the day, the important point to keep in mind is that the deparametrized evolution is simply scale transformations. This underlines the fundamental role played by the CVH Poisson algebra generated by the Hamiltonian constraint and the dilatation generator, which seems to contain the whole dynamics and physical content of the theory.

\subsection{Dirac's observables and  reduced phase space}

Another way to describe the physical content of the theory, which is generically equivalent to deparametrization, is to identify Dirac observables, i.e. constants of motion Poisson-commuting with the Hamiltonian constraint:
\be
\dot{\mathcal{O}} = \{ \mathcal{O} , \cH \} \sim 0\,,
\ee
where we assume that the Hamiltonian constraint vanishes, $\cH=0$. The Poisson algebra of these Dirac observables defines the reduced phase space of physical observables.

Here, we identify the Dirac observables for the cosmological evolution.
%
%
To start with, we already know that the scalar field momentum $\pi_{\phi}$ is a constant of motion, $\{ \pi_{\phi} , \cH\} = 0$, and is a (strong) Dirac observable. But we also know that the dilatation generator is a constant of motion and thus a Dirac observable:
\be
\label{cpi}
\dot{\cC} = \{ \cC, \cH \} = - \cH \sim 0\,.
\ee
Actually these represent the same Dirac observable. Indeed, solving the Hamiltonian constraint yields:
\be
\cC \sim \eps\frac{\pi_{\phi}}{\ka}
\,,\quad
\eps=\pm
\,.
\ee
Moreover, since both $\cC$ and $\pi_{\phi}$ are both constants of motion, the sign $\eps$ is also constant along trajectories and one can not hop from one branch to another along a single trajectory. 

Next, we identify Dirac observables involving the volume $v$ and its conjugate variable $b$:
\be
\cV_{\eps}=v e^{- \eps\ka\phi}
\,,\quad
\cB_{\eps}=b e^{+ \eps\ka\phi}=\f{\cC}{\cV_{\eps}}
\,,
\ee
\be
\{\cV_{\eps},\cH\}
=
\f{\cV}{v}
\left[
\cC- \eps\frac{\pi_{\phi}}{\ka}
\right]
\sim0
\,,\quad
\{\cB_{\eps},\cH\}\sim0
\,,
\ee
where the weak equality only holds on the corresponding branch, i.e. assuming $\pip=\eps\ka\cC$. Let us emphasize that those Poisson brackets do not weakly vanish if we only assume that $\cH=0$. Thus, $\cV_{\eps}$ and $\cB_{\eps}$ can only be thought of as Dirac observables if we choose the $\eps$-branch and switch the original Hamiltonian constraint $\cH=0$ for the stronger constraint $\pip=\eps\ka\cC$. We will see in the section \ref{Obs-so32} how to upgrade them to complete Dirac observables.

Comparing these with the deparametrized trajectories, given in \eqref{vphitraj} and \eqref{deparamtraj}, we see that they are indeed constants of motion. They are the canonical pair of conjugate variables on the reduced phase space:
\be
\{\cB_{\eps},\cV_{\eps}\}=1\,.
\ee

\section{$\SL(2,\R)$ Unitary Representations}
\label{su11irrep}

At the quantum level, the Poisson brackets of the $\sl(2,\R)\sim\su(1,1)$ Lie algebra becomes operator commutators:
\be
[\hJ_{z},\hK_{\pm}]=\pm \hK_{\pm}
\,,\quad
[\hK_{+},\hK_{-}]=-2J_{z}
\,.
\ee
These commutators can be represented  as acting on states $|\mathfrak{C},m\ra$, at fixed value of the Casimir $\mathfrak{C}$, as:
\begin{align}
\hJ_z \,| \mathfrak{C}, m \ra &= m \,| \mathfrak{C}, m \ra \\
\hK_{+} \,| \mathfrak{C}, m \ra &= \sqrt{m(m+1) - \mathfrak{C}}\, | \mathfrak{C}, m +1 \ra \\
\hK_{-}\, | \mathfrak{C}, m \ra &= \sqrt{m(m-1) - \mathfrak{C}}\, | \mathfrak{C}, m - 1 \ra \\
\what{\mathfrak{C}} \,| \mathfrak{C}, m \ra &= \mathfrak{C}\, | \mathfrak{C}, m \ra
\end{align}
The unitary (irreducible) representations of the $\sl(2,\R)$ Lie algebra are obtained by making sure that $\hJ_z^{\dagger}=\hJ_z$ (i.e. keeping $m\in\R$) and $\hK_{+}^\dagger=\hK_{-}$.
This leads to  three classes of unitary representations:
\begin{itemize}
\item {\it The discrete series}:
We distinguish the positive and negative series, which consist respectively in lowest and highest weight representations. They are labelled by half-integers $j\in\f\N2$, $j\ge 1$:
\be
\cD^{+}_j =\bigoplus_{m\in j+\N}\C\,| j, m \ra
\,,\quad
 \cD^{-}_j =\bigoplus_{m\in -j-\N}\C\,| j, m \ra
 \,.
\ee
The Casimir is given by $\mathfrak{C} = j (j-1) \geqslant 0$. These can be constructed from the canonical quantization of the pair of harmonic oscillators in the spinorial formulation discussed in the  section \ref{spinorial}, when realizing the $\sl(2,\R)$ algebra generators as $E$, $K_{z}$ and $L_{z}$, as used for instance in \cite{Livine:2012mh} to define cosmological coherent states.

\item {\it The principal continuous series}: they are labelled by a real positive number $s\in\R^+$ and a parity $\eps=\pm$. Even parity representations are spanned by  states with integer magnetic moment $m\in\Z$ while odd parity representations are spanned by half-integer states $m\in\Z+\f12$:
\be
P^{+}_s= \bigoplus_{m\in \Z}\C\,| j, m \ra
\,,\quad
P^{-}_s= \bigoplus_{m\in \Z+\f12}\C\,| j, m \ra
\ee
The quadratic Casimir is now strictly negative, $\mathfrak{C} = - \left( s^2 + \frac{1}{4} \right) < 0$. It can be written as $\mathfrak{C} = j(j-1)$ with $j=\f12+is$. These are the $\SL(2,\R)$ representations that we use in the present work to quantize the cosmological phase space. 

\item {\it The complementary series}: They are labelled by a real number $\f12<j<1$ and interpolate between the discrete series starting at $j=1$ and the principal continuous series on the imaginary line $\mathrm{Re}(j)=\f12$:
\begin{align}
P^c_j= \bigoplus_{m\in \Z}\C\,| j, m \ra.
\end{align}
The Casimir is given by $\mathfrak{C}=j(j-1)$, with $- 1/4 < \mathfrak{C}  < 0$. These representations do not appear in the Plancherel decomposition for $L^2$ functions on $\SL(2,\R)\sim\SU(1,1)$.

\end{itemize}

\end{document}